# A Historical Method Approach to Teaching Kepler’s 2nd law

Wladimir Lyra, New Mexico State University, USA

**ABSTRACT**

*Kepler’s 2nd law, the law of the areas, is usually taught in passing, between the 1st and the 3rd laws, to be explained “later on” as a consequence of angular momentum conservation. The 1st and 3rd laws receive the bulk of attention; the 1st law because of the paradigm-shift significance in overhauling the previous circular models with epicycles of both Ptolemy and Copernicus, the 3rd because of its convenience to the standard curriculum in having a simple mathematical statement that allows for quantitative homework assignments and exams. In this work I advance a method for teaching the 2nd law that combines the paradigm-shift significance of the 1st and the mathematical proclivity of the 3rd. The approach is rooted in the historical method, indeed, placed in its historical context, Kepler’s 2nd is as revolutionary as the 1st: as the 1st law does away with the epicycle, the 2nd law does away with the equant. This way of teaching the 2nd law also formulates the “time=area” statement quantitatively, in the way of Kepler’s equation,* $M = E - e \sin E$, *(relating mean anomaly M, eccentric anomaly E, and eccentricity e), where the left-hand side is time and the right-hand side is area. In doing so, it naturally paves the way to finishing the module with an active learning computational exercise, for instance, to calculate the timing and location of Mars’ next opposition. This method is partially based on Kepler’s original thought, and should thus best be applied to research-oriented students, such as junior and senior physics/astronomy undergraduates, or graduate students.*

Kepler’s 2nd law is arguably the most challenging of Kepler’s laws to teach. Yu et al. (2010) found that, in a sample of 112 undergraduate student interviews to gauge prior knowledge for an introductory astronomy course, the majority (54%) declined to even guess an answer when inquired about it. This contrasts to the 1st and 3rd laws, where a majority of the same students gave incomplete but correct statements about them (“orbits are not circles”, and “planets orbiting closer to the Sun move faster; those orbiting farther move slower”, respectively). While the sample of Yu et al. (2010) was of nonmajor freshmen, the 2nd law remains underrated in upper division major undergraduate courses, where students’ understanding of it still lingers on the qualitative, and divorced from its historical significance. In addition, Aktan & Dinçer (2014) find alternative conceptions about the 2nd law even among pre-service science teachers. This evidences shortcomings about the way the law is traditionally explained. The 2nd law is frequently taught (e.g. Carroll & Ostlie 2007, Ryden & Peterson 2020) as a variation of the following statement

*A line drawn from the Sun to a planet sweeps out equal areas in equal time intervals.*

This sentence is usually followed by diagrams showing short wide areas near the Sun and long slender areas far away. “They’re equal!”, says the instructor, almost like a curiosity. “Why is this important?”, asks the inquisitive student. Planets go fast near the Sun and slow far from it, is the usual answer. It is how it is described in popular astronomy books (e.g, de Cayeux & Brunier, 1982), in high school physics (e.g. Kuhn 1979, Guimarães & Fonte Boa 2006), and in numerous educational websites. In college, one walks the extra mile of proving from Newton’s laws that Kepler’s 2nd simply reflects angular momentum conservation (e.g. Halliday & Resnick 1960, Marion & Thornton 1988). This is usually done by climbing down the pedestal of differential and integral calculus to the pedestrian world of Euclidian geometry and defining the strange concept of “areal velocity”, which is then proved to be constant.

The historical inversion (Newton before Kepler) is rooted on a discipline approach (Rice 1972). By grounding Kepler’s 2nd law on angular momentum conservation, it draws of principled conceptual knowledge (Leinhardt 1988), facilitating learning by structuring it around a major concept of the discipline of physics. On the other hand, teaching the material in this way has the unfortunate drawback of reducing Kepler’s 2nd law to a post-factum instead of presenting it as the product of original logic, painstaking problem-solving, and what was then cutting-edge research. These in turn, are precisely the skills that should be developed in high-ability students (Dixon et al. 2004). More importantly, this presentation fails to obviate that the 2nd law is about quantitatively finding the planet in the orbit. In this paper I develop a model for teaching Kepler's second law using the historical method (Matthews 1989, Lopes Coelho 2009, Galili 2010), partially recreating in the classroom the historical perspective in which Kepler discovered the law in *Astronomia Nova* (Kepler 1609, Aiton 1969, Boccaletti 2001).

The historical method (also called genetic approach) has a rich record in physics. Over a century ago, Mach (1895, 1911) and Duhem (1906) argued that retracing the original line of thought in discovering the laws of nature led to a deeper understanding of the subject by the novice students. The sentiment is echoed by Schwab (1962), who defined teaching *science as inquiry* in its essence as "to show some of the conclusions of science in the framework of the way they arise and are tested". Modern pedagogy frames this postulate under the idea of cognitive recapitulation (Piaget 1970, Posner 1982): ontogeny recapitulates phylogeny, i.e. there is a parallel between how an individual accrues knowledge (ontogeny) and how the knowledge in the discipline itself evolved (phylogeny). The historical method also highlights the correspondence principle (Bohr 1920), which requires a new theory to explain all the phenomena for which a preceding theory was valid. As such, it prompts the student to understand under what circumstances the previous theory was judged convincing. Equally important, by recreating the atmosphere of discovery, the historical method inherently brings into the classroom the culture of the field (Conant 1964, Holton 1978), informing not only knowledge but also its structure: the deductive logic and reasoning by which knowledge was constructed, older and now obsolete principles and concepts, how they were replaced, the thinking of generations of astronomers -- what Galili (2010) calls the *periphery* of the discipline, as opposed to the discipline's nucleus (axioms and laws) and body (applications). Finally, research in human cognition shows that learning is facilitated when presented with a contrast between alternatives (Waxer & Morton 2012). As a consequence, another reason the historical method is effective as a teaching tool is because, in recreating the narrative that led to the production of knowledge, it also recreates the conceptual conflict that necessitated that knowledge (Monk & Osborne 1997).

This paper is structured as follows. In the next section the context of the study is presented, followed by the teaching design, including a teacher voice's exposition and figures to include in the presentation for replicating the method. I conclude with an assessment of student learning and discussion.

## CONTEXT OF THE STUDY

This method was originally developed as part of a one semester course on dynamical astronomy for physics students in their junior year, at a primarily undergraduate university in California, in 2018. The class had 24 students. I taught it again twice for first-year astronomy PhD students (ASTR 503, "Fundamental Astronomy") at New Mexico State University, in 2019, and 2020. I teach this in two classes of 75 minutes each, as part of a module on Kepler's laws. The first two times were taught in person, the third time online in "flipped classroom" format (King 1993; the videos are available at this url). The 2019 class had 10 students, which provided a convenience sample (Saumure & Given 2008) to poll about the 2nd law.

The fundamental question that guides the module is a question that intrigued humanity for millenia: *how to predict the position of the planets?* The goal of the module is to understand how Kepler's laws connect to the emergence of modern astronomy, to understand planetary motion, the interplay between theory and observations, and the fundamental importance of observational accuracy. The module on Kepler's laws is done after a module on Spherical Astronomy, so the students are familiar with coordinate systems on the celestial sphere, and how to transform between equatorial and ecliptic coordinates. I also introduce the concept of elongation, the angle between the planet and the Sun. This serves the purpose of introducing Ptolemy's model, which is key to understanding the revolutionary character of Kepler's 2nd law.

For the junior class I started with Owen Gingerich celebrated Mars' lab (Gingerich, 1983) to find Kepler's 1st law in an active learning way (Bonwell & Eison 1991). For the PhD course Gingerich's lab was done as a computational exercise. The lab is followed by the geometrical proof that the orbit is an ellipse, again using the historical method, with pre-Newtonian reasoning (Appendix A). In the online version I could not provide drafting tools to each student, so I created a video of the method (available at this url), and we did active learning in class showing how we would not be able to discriminate between an ellipse and an off-centered circle with the accuracy of common classroom drafting instruments. The 1st law lab and instruction set the stage for Kepler's 2nd law.

## TEACHING DESIGN

The historical method was adopted in response to the unsettledness of teaching Kepler's 2nd Law as a post-factum, and with seemingly less importance than the 1st and the 3rd. The 1st law, *the planets orbit the Sun in elliptical orbits with the Sun at one of the two foci*, has a clear and powerful paradigm-shifting formulation. Its statement is a direct and unequivocal breaking with the previous cosmological models, of both Ptolemy and Copernicus, that insisted on circular orbits. The 3rd law, *the cube of the semimajor axis is proportional to the square of the periods*, is formulated as an elementary mathematical statement, and thus conveniently translated into quantitative homework assignments and exams, even at the high school level. In contrast with the 1st and the 3rd, the second law, *the radius vector connecting the planet to the Sun sweeps equal areas in equal times*, sounds disturbingly turbid to the modern student, its geometric statement a remnant of a pre-calculus era. Transplanting ourselves to Kepler's time by putting aside Newtonian physics and knowledge of conservation of angular momentum, one should ask: *why did Kepler care about area*? In the junior class in 2018 I asked the students this question. No hands were raised. Not wanting to repeat the usual way of teaching the 2nd law, and given how other educators also struggle with how to present it (Setyadin et al. 2020), I decided to teach it partially following the historical method to retrace Kepler's original line of thought.

In its historical context, Kepler 2nd law is similar in formulation to the 1st law in the sense that it is contrasted to the previous model. *The orbit is an ellipse* contrasts to *the orbit is a circle*. Likewise, *equal areas at equal times* contrasts to *equal angles at equal times*. The 2nd law is formulated as a conceptual conflict. For 1500 years, up to Kepler, astronomy insisted not only on circles but also on uniform motion. In Ptolemy's model, to account for the perceived non-uniform motion of a planet, he introduced the equant, which is a point on the line of apsides about which the center of the epicycle does uniform motion. The practicality is that time is given by angle, so the motion of the planet, though non-uniform from Earth's reference frame, is easily parametrized in time. The doctrine of uniform motion was so prevalent that it was an *a priori* in Copernicus new heliocentric model: to do away with non-uniform motion and cast all planetary movement as uniform circular motion in deferents and epicycles about their centers. Every astronomer up to Kepler had it ingrained that there was a reference frame about which a planet sweeps equal angles at equal times. That is the prior model that the 2nd law contests. Imagine that you never heard about circular orbits before. It would become difficult to understand the paradigm-shifting impact of the 1st law. That is precisely the situation a modern student encounters the 2nd law.

The remainder of this section will be presented in "teacher's voice", as if in the classroom. The method here developed to teach Kepler's 2nd law shares similarities with the presentation of the same subject by Holton & Brush (1952). The method, however, is more mathematically grounded and focuses on the correspondence principle and conceptual conflict between the equant model and Kepler's 2nd law. A drawback of the method is the need to teach the equant model, which many students (as well as instructors) have little familiarity with. However, we minimized the time needed to introduce the model, while also keeping it pedagogical.

## Teacher's voice presentation

Finding the shape of the orbit is not solving the whole problem of planetary motion. A practical question remains: how to find the planet in the orbit? To better understand Kepler's 2nd law, let us look at what existed before him. Ancient wisdom insisted in uniform circular motion, because it was their way to understand periodicity. Regularity was found in circles, according to Copernicus (1543), "the only figure that can bring back the past". Although it was apparent that the Sun (or the Earth for that matter) was not the center of the orbit, the shape of the ellipse was out of the reach of their observational accuracy. Non-uniform motion along the orbit was also evident, and a solution was found by Ptolemy, namely, the equant model. Most modern astronomers are not familiar with this model, which is nowadays only of historical importance, so a brief pedagogical exposition is warranted. Let us get to it step by step.

### Starting simple: a sun-centered, circular model

Let us assume that planets go in circular orbits, centered at the Sun, in uniform motion (Fig. 1a). At any instant of time,

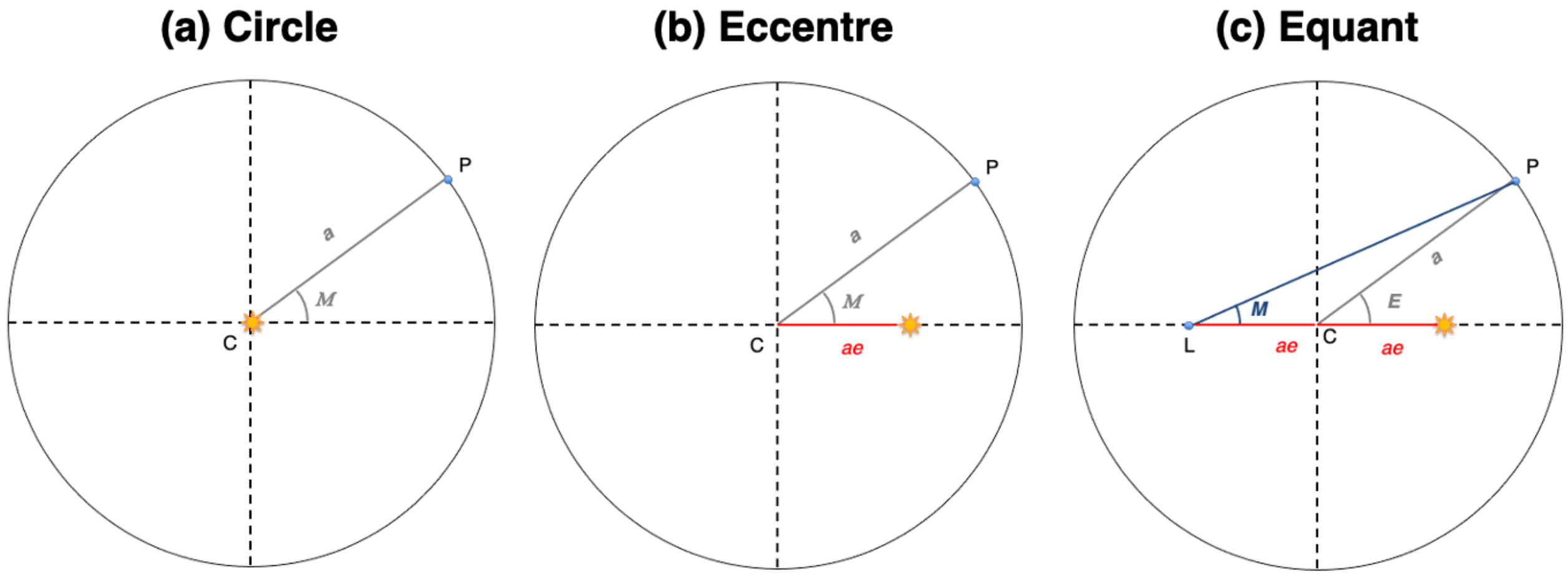

**Figure 1:** Elements of the different circular orbit models: (a) Sun-centered uniform circular motion about the Sun; (b) Off-centered uniform circular motion about the center; (c) Off-centered uniform motion about the equant (point L). $M$ is the mean anomaly and $E$ the eccentric anomaly. On (a) and (b) the mean anomaly and the eccentric anomaly are identical.

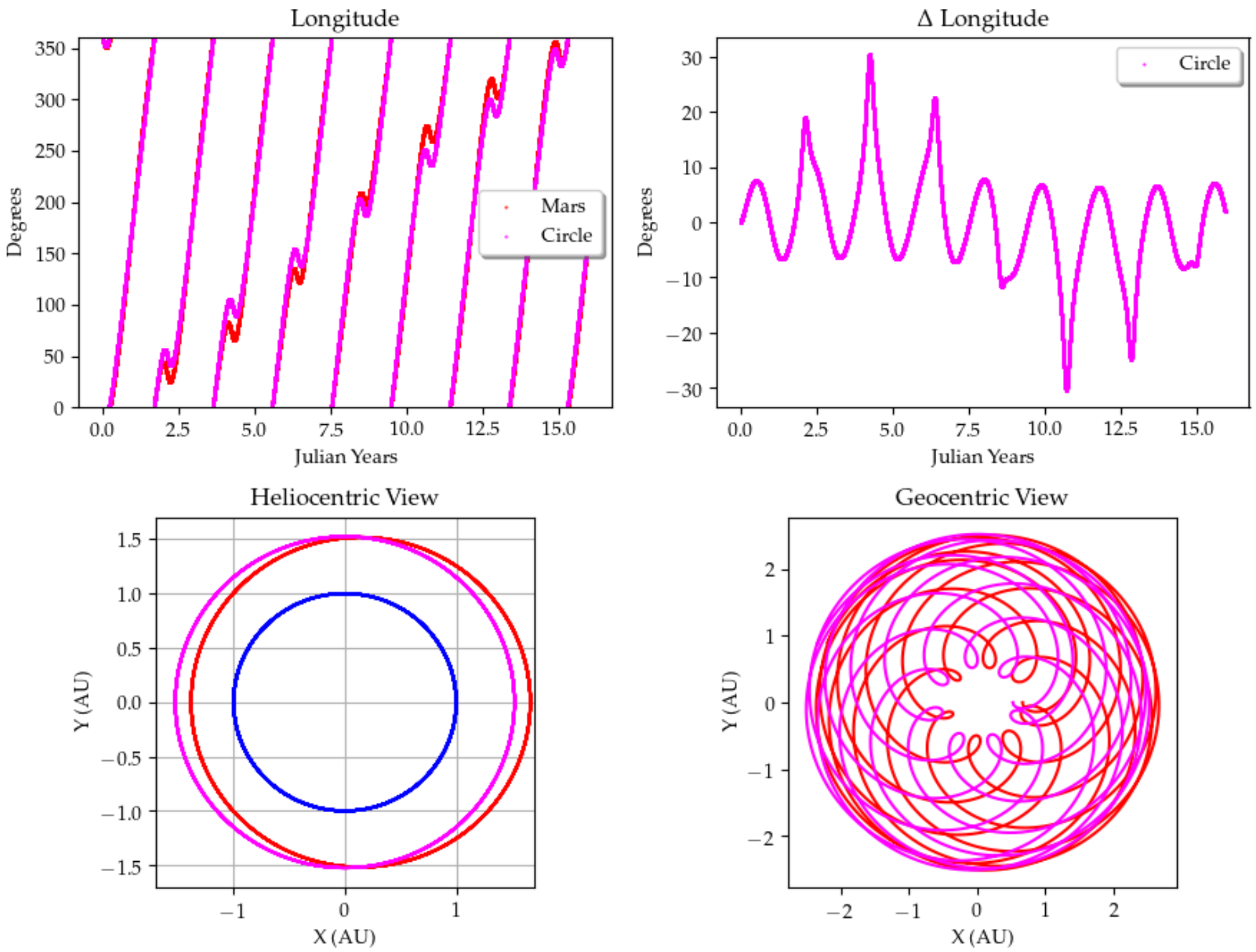

**Figure 2:** Mars orbit versus a circular Sun-centered orbit model. Upper left plot: Ecliptic longitude vs Time, Mars (red) vs circular orbit (magenta). The longitudes generally match, except at retrogradations. Upper right plot: Longitude residual. The error amounts to as much as 30 degrees. The circular Sun-centered model is not acceptable. Lower left plot: Heliocentric view of the orbit. Red is mars, blue is Earth, magenta the model. Lower right plot: Geocentric view of the orbit.

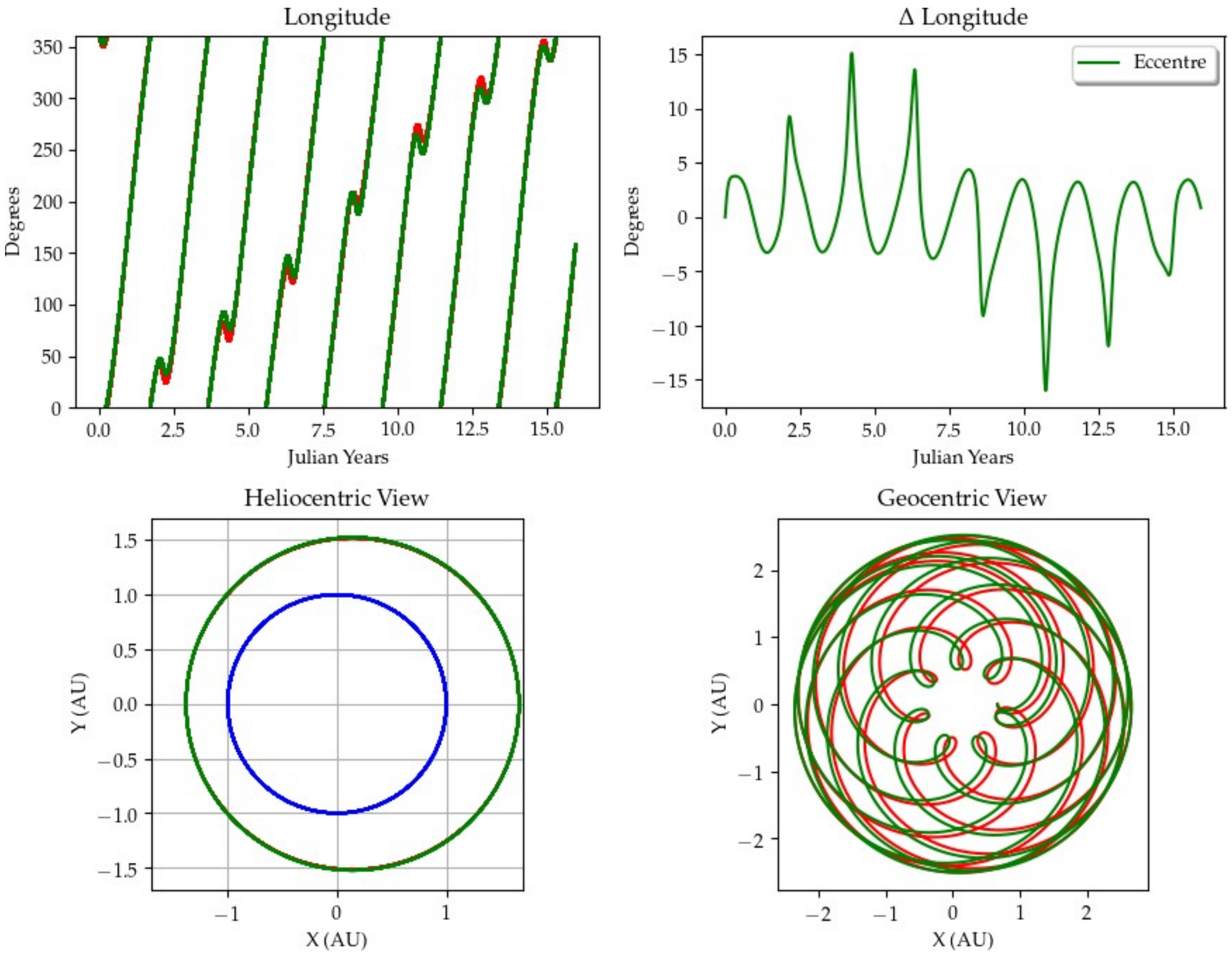

**Figure 3**: Mars orbit versus off-centered circular orbit model (the eccentre), keeping uniform motion. Upper left plot: Ecliptic longitude vs Time, Mars (red) vs circular off-centered orbit (green). The longitudes generally match, except at retrogradations. Upper right plot: Longitude residual. The error is better than the Sun-centered model, but still amounts to as much as 15 degrees. The off-centered model with uniform circular motion is not acceptable. Lower left plot: Heliocentric view of the orbit. Red is mars, blue is Earth, green the model. Lower right plot: Geocentric view of the orbit.

the position of Mars is given by

$$x(t) = a \cos M(t) \quad (1)$$

$$y(t) = a \sin M(t) \quad (2)$$

where $M(t) = nt$, with $t$ meaning time, and $n = 2\pi/T$ is the mean motion, where $T$ is the period of Mars. Here, $M$, the mean anomaly, is equal to both the eccentric and true anomalies. Notice that everything but time in the definition of mean anomaly is constant. Mean anomaly equals time. *Mean anomaly is time.*

The four panels in Fig. 2 show (upper left plot): the ecliptic longitude vs time for this model (circular and sun-centred orbit, in magenta) and the actual position of Mars, in red; (upper right plot): deviation between model and actual Mars; (lower left plot): The bird-eye heliocentric view of the orbit; (lower right plot): the geocentric view of the orbit.

The model does not reproduce either the shape of the orbit or the longitudes. The predicted positions of retrogradations, specifically, are off by as much as 30 degrees from the actual positions of Mars. The model has to be discarded.

**The Eccentre**

The next model is the eccentre (Fig 1b). This model merely shifts the position of the center of the orbit away from the Sun by an amount $ae$, keeping the uniform motion. At any given instant in time, the position of Mars, seen from the

Sun, is now given by

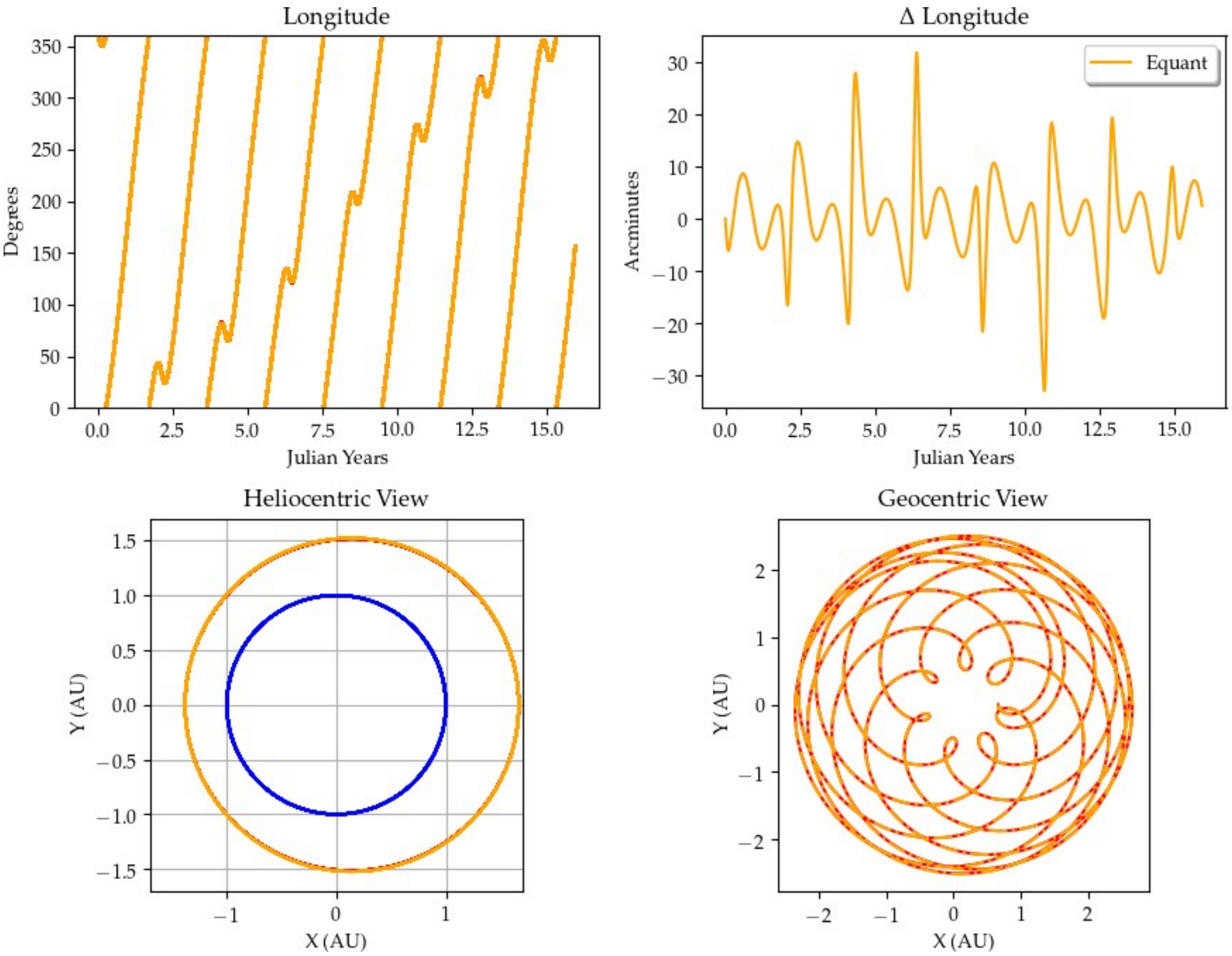

**Figure 4**: Mars orbit versus off-centered circular orbit model, with uniform motion about the equant. Upper left plot: Ecliptic longitude vs time, Mars (red) vs equant model (orange). Upper right plot: Longitude residual. The error is at most 30 arc minutes. Ptolemy's accuracy was 1 degree. The model is acceptable. Lower left plot: Heliocentric view of the orbit. Red is mars, blue is Earth, orange the equant model. Lower right plot: Geocentric view of the orbit. The equant model reproduces location and time of retrogradations.

$$x(t) = a \cos E(t) - ae \tag{3}$$
$$y(t) = a \sin E(t) \tag{4}$$

where $ae$ is the amount we shift the center away from the Sun. The angle $E$, the eccentric anomaly, is $E(t) = nt$ and equal to the mean anomaly $M$.

This model is shown in green in Fig. 3. It reproduces the orbit, but does not reproduce the velocity of Mars. It predicts oppositions and retrogradations still off by 15 degrees. Again, this model cannot be right.

**Non-uniform motion**

To fit the velocity of the planets, Ptolemy added a third device, the equant point, defined as a point on the line of apsides about which the angular velocity of a body on its orbit is constant. This point is point L in Fig. 1c. About L, the planet, located at P, goes around in uniform motion, being described by the angle $M = nt$. The eccentric anomaly $E$ seen from the center of the orbit is related to $M$ by noticing that the triangle ΔLCP has angles $L\hat{C}P = 180^{\circ} - E$, and $L\hat{P}C = M - E$. The side CP has length equal to $a$, and the side LC has length equal to $ae$. Applying the law of sines

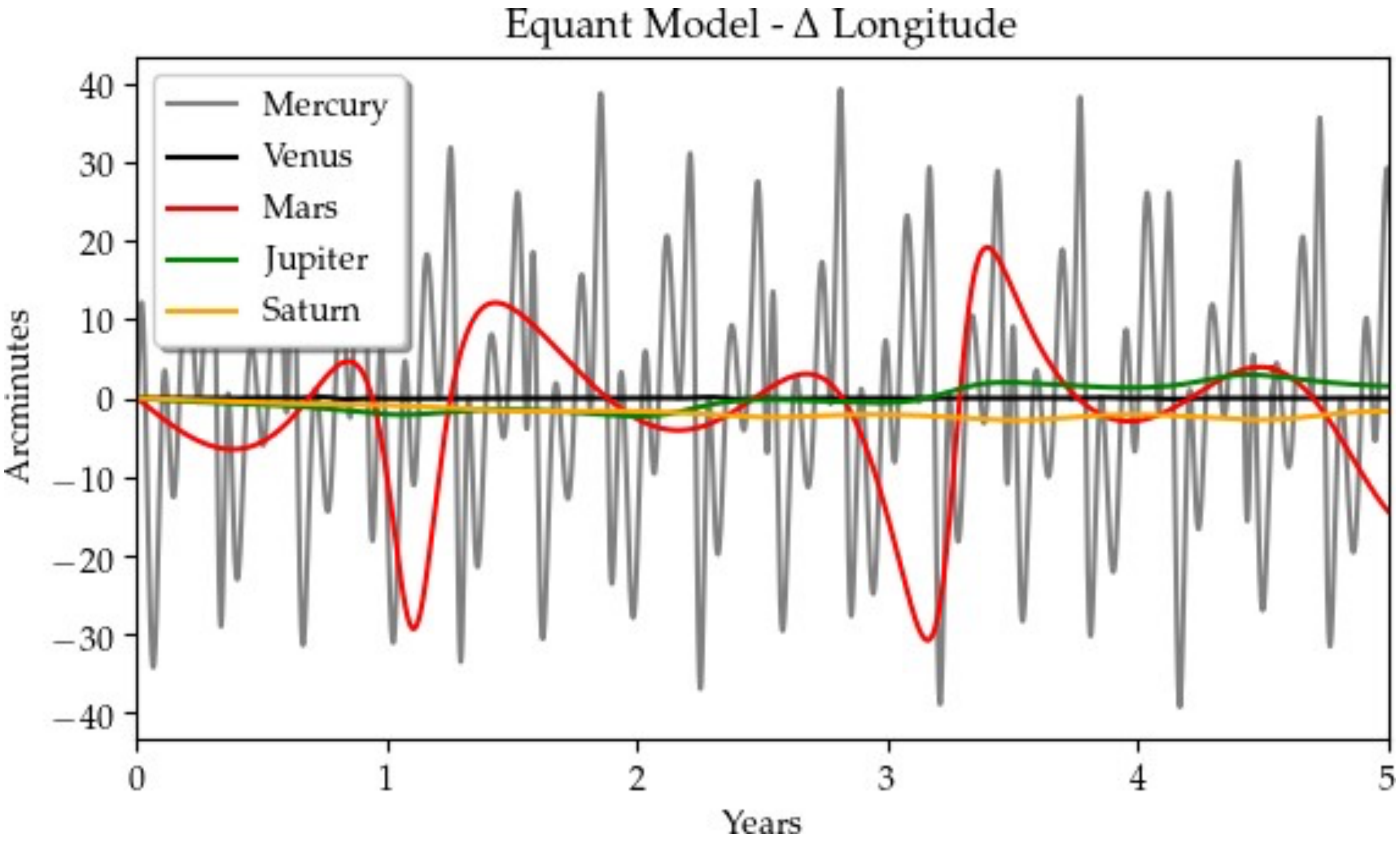

**Figure 5**: Residuals of the equant model for each planet. The residuals reflect orbital eccentricity. Even for Mercury, the most eccentric planet ($e$ = 0.2), the equant agrees with the observations down to 40 arcminutes.

$$\frac{\sin(M-E)}{ae} = \frac{\sin M}{a} \tag{5}$$

that is,

$$\sin(M-E) = e \sin M \tag{6}$$

Thus, solving for $E$

$$E = M - \sin^{-1}(e \sin M) \tag{7}$$

At any time, the position of Mars, seen from the Sun, is again given by Eqs (3) and (4), except that now $E(t)$ is non-uniform. The result is shown in Fig. 4. Agreement is achieved to within half a degree. Considering that Ptolemy did not have accuracy under a degree (Høg 2017), the equant gives *excellent* agreement to the observations. Fig. 5 extends the equant model to other planets. Each of them has their own equant – which simply reflects the eccentricity of the orbit. Even in the case of Mercury, the planet of highest eccentricity, the agreement with the observations is satisfactory to the degree. This figure also shows how appropriate Mars was as subject of Kepler's analysis. Of the superior planets, it is the one whose deviation more blatantly disagreed with the prevailing model. Venus, Jupiter, and Saturn deviate by less than 2 arcminutes, within the accuracy of Tycho's data (Høg 2017). Mercury, never too far from the Sun, is simply too difficult to observe.

**Optional**

The instructor may want at this point to do a parenthetical comment, returning for a moment to modern scientific parlance, and noting that the equant model is a model accurate to first order in eccentricity (Hoyle 1973, Murray & Dermott 1999). At higher eccentricities the equant model will again start to deviate significantly from the observations. Fig. 6 shows a hypothetical planet of eccentricity $e$ = 0.45. The equant model is off by more than 10 degrees. While for $e \sim 0.2$, like Mercury's orbit, the model is satisfactory down to 40 arcmin accuracy, for $e$ = 0.45 one would need higher order corrections. Kepler's 2$^{nd}$ law is the full solution.

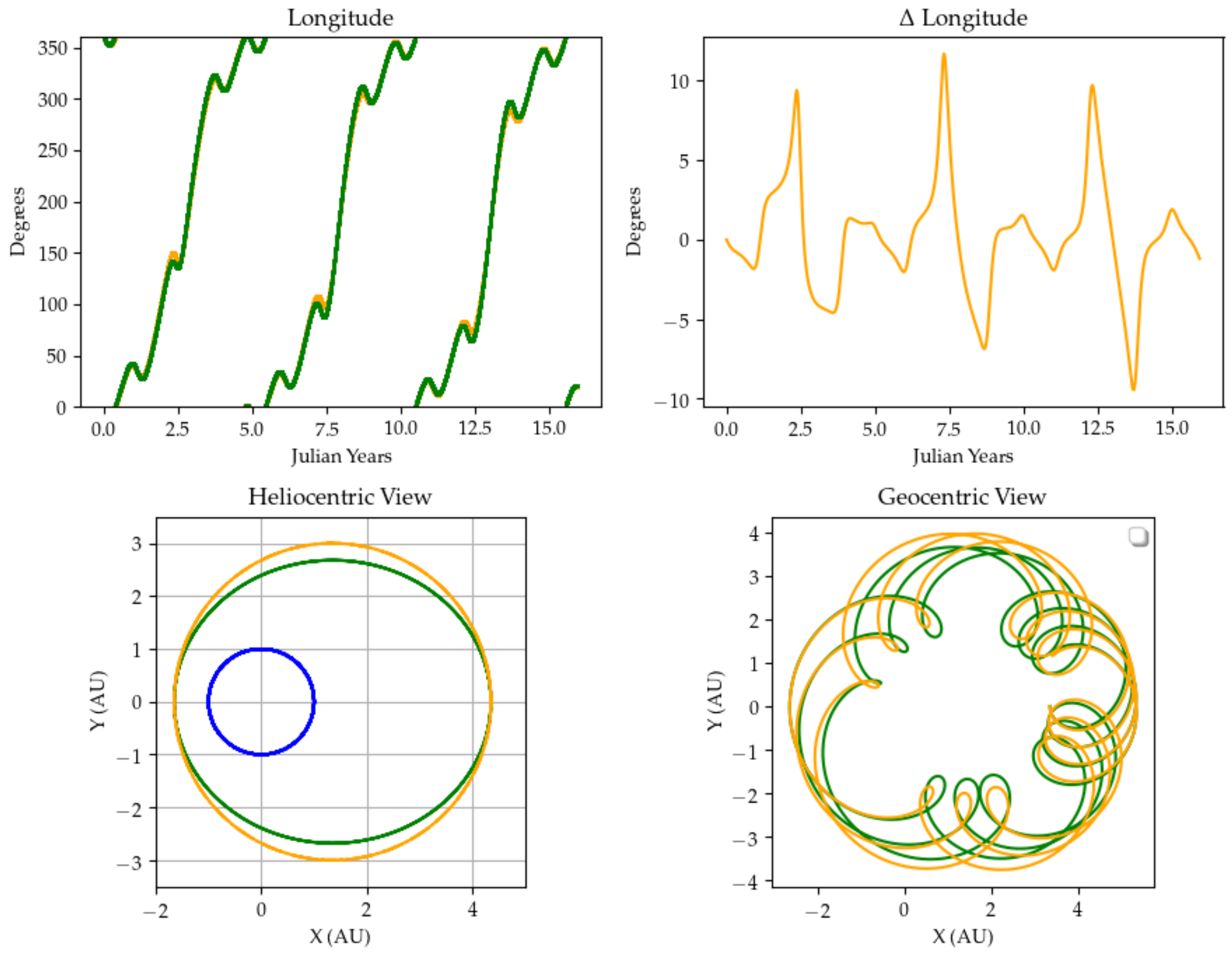

**Fig 6.** Validity of the equant model. A hypothetical planet (green) with orbital eccentricity 0.45, The equant model (orange) does not reproduce it well anymore. The equant is a model accurate only to 1st order in eccentricity.

Ptolemy's solution had a very practical function. Given a point in the line of apsides upon which the planet sweeps equal angles in equal times, the orbit can be parametrized, as given by Eqs. (3) and (4) with $E$ given by Eq. (7). Kepler had two problems. First, Tycho's observations of Mars, accurate to 2 arcmins, did not allow for the 30 arcmin error given by the equant model. Second, his ellipses, with the planet speeding nearing perihelion and slowing down nearing aphelion begged the question: *what is the equivalent to the equant?* What is the point about which a planet sweeps equal angles in equal times? Where is the point along the line of apsides that we can say that angle equals time? As it turns out, Kepler's quest to answer this question culminated with his 2nd law, that demolished the idea of uniform motion. The answer is: *there is no equant*. For the ellipse, there is no point about which an observer will see equal angles at equal times. *Kepler's first law does away with the epicycle. Kepler's second law does away with the equant.*

**The ellipse has no equant**

This subsection is the one directly from Kepler's *Astronomy Nova*. I taught it in detail twice, before realizing that the level of detail is unnecessary. What the students should know is that ***Kepler tried to find the equant for the ellipse and failed to find it***. In trying to find out the location of Mars' equant, Kepler again made use of Tycho's data. He took four observations of Mars in opposition, at times $t_1$, $t_2$, $t_3$, and $t_4$, corresponding to mean anomalies $M_1$, $M_2$ $M_3$, and $M_4$. The equant would be the unique point on the line of apsides whence Mars is seen at these angles (Fig. 7).

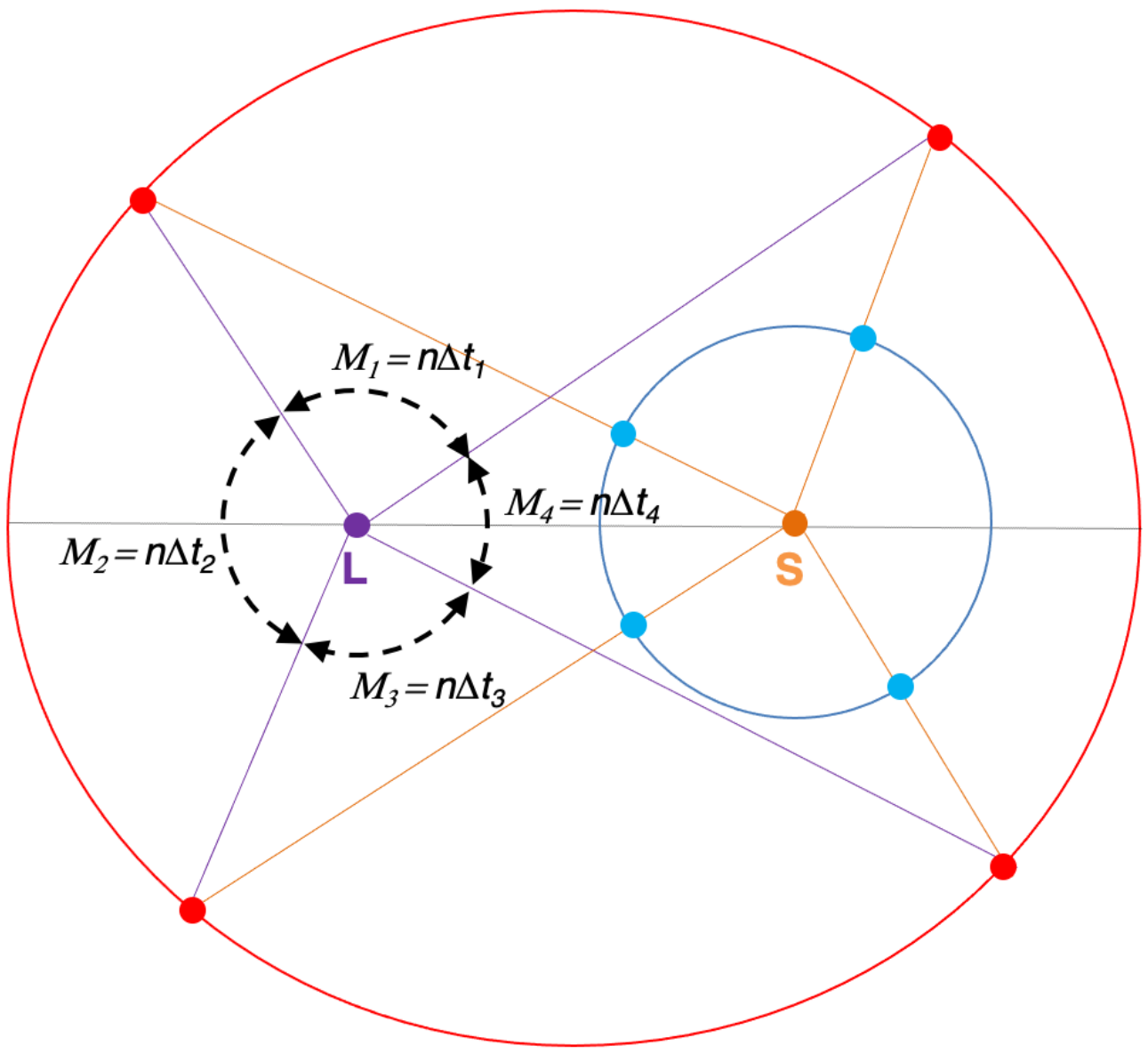

**Fig 7.** Kepler's method to determine the location of Mars' equant in an elliptic orbit, using four observations of Mars in opposition. The four dots in each orbit represent the four observations; the motion is counterclockwise. Because the Sun (S), Earth (blue orbit and dots), and Mars (red orbit and dots) are aligned, the orange lines intersect at the Sun. Timing the observations yields the mean anomalies $M_1$, $M_2$, $M_3$, $M_4$. Kepler then looked for the point L in the line of apsides where Mars would be seen at exactly these angles, failing to find it. The model could not be reconciled with the observations by 8 arcminutes, inadmissible by Tycho's 2 arcminute accuracy, forcing him to discard the equant model. The eccentricity of Mars' orbit is highly exagerated (~0.4 instead of ~0.1) for clarity.

Yet, Kepler's best fit with an "ellipse equant" model was incompatible with the observations by 8 arcminutes, which was inadmissible by Tycho's 2 arcminute accuracy. Kepler had to go back and question his assumptions. But the assumptions were minimal. They amounted to

1) Mars orbits the Sun;

2) Tycho's observations are reliable;

3) The equant exists.

(1) and (2) were beyond doubt correct. The conclusion was astonishing. The equant, a staple of astronomy for 1500 years, cannot exist. Kepler started this analysis by asking the question: where is the equant? And the answer was: there is no equant. There is no point about which we can say the planet sweeps equal angles at equal times. Uniform motion does not exist. Time is not given by angle. This is the aspect of the second law that should be emphasized: it rules out 1500 years of the paradigm of uniform motion.

**Time is equal area**

Kepler had disproved fifteen centuries of "equal angles at equal times". That still leaves the problem of how to find the eccentric anomaly as a function of time. In looking for something that could be a measurement of time, Kepler stumbled on what this something was.

Because planets are slower when far from the Sun and faster when close, Kepler reasoned that the velocity was inversely proportional to the distance, $u \propto 1/r$. If that is the case, one can multiply both sides by time and write the proportionality

$$u\, r\, t \propto t \tag{8}$$

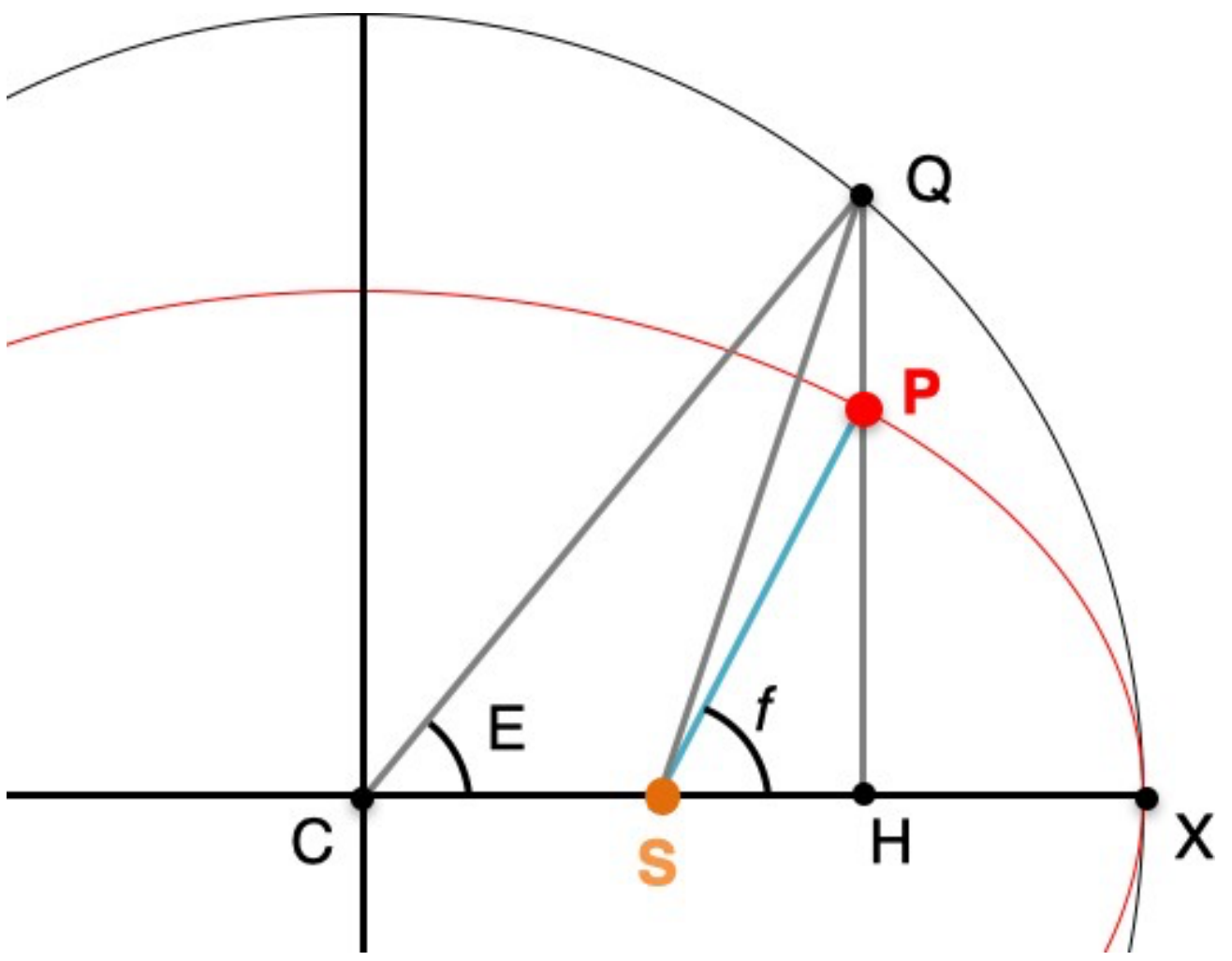

**Figure 8:** The true anomaly $f$ is the angle, with vertex at the Sun, from perihelion to the planet. The eccentric anomaly $E$ is the angle with vertex at the center of the orbit, from perihelion to the planet. Given the geometry of the ellipse, PH/QH = $b/a$, where $b$ is the semiminor axis and $a$ is the semimajor axis.

The quantity in the left-hand side, whatever it is, is linearly proportional to time. It is the "something" sought in equal "something" at equal times. But what is its interpretation?

The product $ut$ is the length of the arc swung by the planet. If the time is infinitesimal, $t \rightarrow dt$,the arc is infinitesimal, $dl = udt$. Then the Sun, the planet's position at $t$ and its position at $t+dt$ form a triangle, of area $r\,dl\,/\,2$. Comparing to Eq. (8), the quantity $u\,r\,dt = r\,dl$ that is linearly proportional to time is thus the *area.* As a planet orbits the Sun, the area it sweeps is proportional to time. Mean anomaly is not given by an angle. Mean anomaly is given by an area.

**Kepler's Equation**

At this point the class becomes finding the mathematical statement of Kepler's equation. The students have already studied in the 1st law the geometry of the ellipse (Appendix A).

The mean anomaly is proportional to the area. The question then is, how to compute the area? Kepler did not know calculus, so he could not calculate the area by summing the infinite infinitesimal distances. But Kepler was an excellent geometer. After he discovered that the orbit was an ellipse, he used the geometry of the ellipse to find out the area. Most students do not know a proof that the area of the ellipse of semimajor axis $a$ and semiminor axis $b$ is $\pi ab$, so a short proof of it using integration is shown in Appendix B.

Given that the planet sweeps equal areas at equal times, a relation between mean anomaly (time) and area can be established. A planet sweeping equal areas at equal times will, within a time $t$, sweep an area

$$A_{\text{sector}} = \pi ab \frac{t}{T} \tag{9}$$

where $T$ is the orbital period. The question is, what is the area of the sector? Consider Fig. 8. The area swept from perihelion (X) to point P is the area of the elliptic sector $A_{SPX}$. We can relate it to the area of the circular sector $A_{SQX}$ by using the relation PH/QH = $b/a$.

$$A_{\text{SPX}} = \frac{b}{a} A_{\text{SQX}} \tag{10}$$

The elliptic sector $SQX$ can be broken down as the circular sector $CQX$, minus the triangle $\Delta CSQ$

$$A_{\text{SQX}} = A_{\text{CQX}} - A_{\text{CSQ}} \tag{11}$$

The circular sector *CQX* comprises an angle *E* of the full $2\pi$circle, so its area is

$$A_{\mathrm{CQX}} = \pi a^2 \frac{E}{2\pi} \tag{12}$$

As for the triangle $\Delta CSQ$, its base is $CS = ae$, and height $QH = a \sin E$

$$A_{\mathrm{CSQ}} = \frac{1}{2}(ae)(a \sin E) \tag{13}$$

We thus find the area of the elliptic sector *SPX*,

$$A_{\mathrm{SPX}} = \frac{ab}{2}(E - e \sin E) \tag{14}$$

But because area equal time, $A_{\mathrm{SPX}} = \pi abt/T$. Equating both,

$$2\pi \frac{t}{T} = E - e \sin E \tag{15}$$

Given $n = 2\pi/T$, the left hand side is $M = nt$, the mean anomaly. Thus,

$$M = E - e \sin E \tag{16}$$

This result is known as **Kepler's equation**. It is a direct consequence of Kepler's 2nd law, and can also be seen as Kepler's 2nd law itself. The left-hand side is time. The right-hand side is area.

After teaching Kepler's 2nd this way, I gave as homework assignment a computational exercise where students had to predict the timing and location of Mars' next opposition, given the orbital elements and the current position of Mars and the Sun. I tested the method in both traditional and flipped classroom (King 1993) environments. In the latter, the lab was started in an online class (taught during the 2020 pandemic of covid-19), with the students sharing a python jupyter notebook via simultaneous video conferencing while freely debating. The method is thus in line with active learning (Bonwell & Eison 1991) and constructivist learning theory (Simon 1995).

## ASSESSMENT OF STUDENT LEARNING

A pre-and-post test was used in the graduate class administrated in 2019; the class had 10 students. The pre-test questionnaire had questions pertinent to all material covered in Fundamental Astronomy, among which one of the questions was "What is the relevance of area in Kepler's 2nd law?". The post-test was administrated after both Kepler's Laws and Celestial Mechanics modules, 3 weeks after Kepler's 2nd law was taught. The table with pre-and-post test answers is shown in Table 1. As seen from the table, all students had the qualitative understanding of the law, as expected from graduate students in a major astronomy research institution, yet no one tied it to the crucial fact of finding the true anomaly. Post-instruction, the answers varied little from the pre-test, with 70% of the students repeating the "equal areas equal times" response of the pre-test. 30% showed a different answer: that area "allows position of the planet to be determined", "is time", and "shows there is no reference point from which an object appears to orbit equal angles in equal times". Perhaps the question of the pre-and-post test could have been better phrased. In contrast, anecdotal feedback I received so far, as well as the degree of student comprehension on dealing with Kepler's equation, supports the approach. One student in particular approached me to say they found the lab "inspiring" and "really interesting", adding in particular that they liked the way I taught the class, going through the historical development of the ideas and the critical thinking involved. One student, after the Mars opposition computational exercise, stated that "now I finally understand Kepler's 2nd law and how to use it". Another student said they liked how the class was taught "like research". Yet another student declared "mad respect" for Kepler after the module.

**Table 1**. Students pre- and post-test answer to the question "What is the relevance of area to Kepler's 2nd law". The post-test was administrated 3 weeks past instruction, and after the next module, on Celestial Mechanics, was taught.

| Pre-Test answer | Responses *N* | Post-Test answer | Responses *N* |
|---|---|---|---|
| Equal areas at equal times | 10 | Equal areas at equal times | 7 |
| | | Determines the planet velocity | 1 |
| | | Allows the position to be determined | 1 |
| | | No equal angles at equal times | 1 |
| | | Conservation of angular momentum | 1 |

## CONCLUSION

In this work I constructed a way to teach Kepler's 2nd law based on the historical method. The perceived benefits of the approach are enumerated below.

1. It frames the teaching in terms of a conceptual conflict, which research in human cognition shows is conducive to more effective learning. The conflict is between the equant model (equal angles) and Kepler's 2nd law (equal areas).
2. The method uses the correspondence principle, making the students understand the validity and limits of the equant model in its own historical framework, as a valid model that reproduces the observations up to about half a degree. The students understand why it worked (a model accurate to first order in eccentricity) and why it had to be discarded (when observational accuracy became better than the 40 arcminute accuracy given by the model). Data of worse quality would not have been able to discern between the equant model and Kepler's 2nd law. As such, the approach also emphasizes to the students the paramount importance of observational accuracy.
3. The proposed approach highlights the revolutionary character of Kepler's 2nd law: instead of repeating "equal areas in equal times" instructors can instead say "contrary to 1500 years of astronomical lore, there is no such thing as equal angles in equal times. Kepler's 1st law discards the epicycle. Kepler's 2nd law discards the equant. Area is how you measure time and hence how you find the planet."
4. The approach is also rooted in mathematical grounding, as "time = area" is stated not only through geometrical illustrations, but by Kepler's equation (Eq. 16). Students can then manipulate it quantitatively, as usually done for the 3rd law.
5. By recreating the atmosphere of discovery, the method also frames the class in terms of cultural teaching, bringing into the classroom the culture of astronomy. It is a narrative method that reveals the inner workings of the minds of the pioneers of the discipline, allowing their own voice to be brought into the classroom. For instance, Kepler famously wrote, in trying to locate the equant (Fig 7): "If this wearisome method has filled you with loathing, it should more properly fill you with compassion for me as I have gone through it at least seventy times at the expense of a great deal of time." That is a feeling that many a graduate student can empathize with. Combining human reason and emotion with the timeless elements of paradigm-shifting research, namely, tension between theory and new data, new data leading to a new theory, the new theory corresponding to the previous theory in its limit of applicability, this is a method that humanizes science and creates a bridge between a student experience and that of the greatest names in the history of the field.

Finally, on the limitations of the method, it has been said that the inquiry method, of which the historical method is a subset, is too difficult for any but the brightest students and that by teaching discarded ideas it is prone to causing confusion (Welsh et al. 1981). Indeed, the method has been tested in upper division undergraduate and graduate studies only, so its effectiveness at lower division or general education courses is unconstrained. Also, because Kepler's equation is transcendental, this method is best used in graduate curricula, where computational techniques are more routinely applied. Another criticism is that scientists are not historians and, by attempting to teach history of science we incur into the danger of teaching bad history (Matthews 1989). Also, as stated by Ausubel (1968) "the most important factor influencing learning is what the learner already knows". Indeed, Table 1 suggests resilience of previous conception in that 70% of students repeated the pre- and post-test answer. The post-test was given after 3 weeks of instruction and after teaching Celestial Mechanics. It is unclear if the time elapsed, the introduction of Newtonian physics, or the phrasing of the question influenced this result.

**ACKNOWLEDMENTS**

I would like to acknowledge productive conversations with Joshua Tan, Mordecai Mac Low, and Roberto Pimentel. A first draft of this paper greatly benefited from comments by Anna Danielsson.

**AUTHOR BIOGRAPHY**

**Wladimir Lyra** is an Assistant Professor of Astronomy at New Mexico State University whose scholarship focuses on planet formation. E-mail: wlyra@nmsu.edu (corresponding author).

**REFERENCES**

Aiton, E. J. (1969). Kepler's Second Law of Planetary Motion. Isis, 60, 75-90.

Aktan, D. Ç. & Dinçer, E. O. (2014). Examination of pre-service science teacher's understanding level of Kepler's laws with ranking task questions. Journal of Baltic Science Education, Vol. 13, No. 2, 276-288.

Ausubel, D. P. (1968). Educational Psychology: A Cognitive View. New York: Holt, Rinehart & Winston, 1968

Bocalleti, D. (2001). From the epicycles of the Greeks to Kepler's ellipse: the breakdown of the circle paradigm. In Cosmology Through Time Ancient and Modern Cosmology in the Mediterranean Area; Monte Porzio Catone (Rome), Italy, June 18-20, 2001.

Bohr, N. (1920), On the spectral series of the elements. Zeitschrift fur Physik, 2, 423-469.

Bonwell, Charles C., and James A. Eison. 1991. Active Learning; Creating Excitement in the Classroom. ASHE-ERIC Higher Education Report No. 1. Washington, D.C.: The George Washington University, School of Education and Human Development.

Carroll, B. W., & Ostlie, D. A. (2007). An introduction to modern astrophysics.

Conant, J. B. (1964). Case studies in experimental science. Harvard university press. Cambridge MA.

Copernicus, N. (1543). On the Revolution of the Heavenly Spheres.

Dixon, F.A. Prater, K. A. Vine, H.M., Wark, M. J., Williams, T., Hanchon, T. & Shobe, C, (2004). Teaching to Their Thinking: A Strategy to Meet the Critical-Thinking Needs of Gifted Students. Journal for the Education of the Gifted. 28, 56–76.

Duhem, P. (1906/1954). The Aim and Structure of Physical Theory. Princeton University Press, Princeton NJ.

Galili, I. (2010). History of Physics as a tool for teaching. In M. Vicentini & E. Sassi (eds.), *Connecting research in Physics Education with Teachers Education.* I.C.P.E. book, pp. 153-166

Gingerich, O. (1983). Laboratory Exercises in Astronomy – The Orbit of Mars. *Sky & Telescope,* 66, 300.

Holton, G. & Brush, S. (1952). Introduction to concepts and theories in physical science. (3rd edition: Holton, G., Brush, S. & Evans, J. (2001). Physics, the Human Adventure: From Copernicus to Einstein and Beyond. Physics Today - 54. 10.1063/1.1420555.)

Holton, G. (1978). On the educational philosophy of the Project Physics Course. In his The Scientific Imagination: Case Studies. Cambridge: Cambridge University Press.

Hoyle, F. (1973). Nicolaus Copernicus: an essay on his life and work. Heinemann, London; 1St Edition (January 1, 1973)

Høg, E. (2017). Astrometric accuracy during the past 2000 years. arXiv e-prints, arXiv:1707.01020.

Kepler, J. (1609), Astronomia Nova.

King, A. (1993). From sage on the stage to guide on the side. College Teaching, 41, 30-35.

Leinhardt, G. (1988). Getting to know: Tracing students' mathematical knowledge from intuition to competence. Educational Psychologist, 23(2), 119-144.

Lopes Coelho, R. On the Concept of Energy: How Understanding its History can Improve Physics Teaching. Sci & Educ 18, 961–983 (2009). https://doi.org/10.1007/s11191-007-9128-0

Mach, E. (1895). On instruction in the classics & the sciences. In Popular Scientific Lectures. Chicago: Open Court. La Salle (1943)

Mach, E. (1911). The history & root of the principle of the conservation of energy. Chicago: Open Court. (orig. 1872).

Matthews, M . R. (1989). A role for history and philosophy in science teaching. Interchange, 20, 2-15.

Monk M. & Osborne J. (1997). 'Placing the History and Philosophy. of Science on the Curriculum: A Model for the Development of Pedagogy.' *Science Education*, 81(4), 405-424

Murray, C. D. & Dermott, S. F. (1999). Solar System Dynamics. Cambridge University Press.

Posner, G. J., Strike, K. A., Hewson, P. W. & Gertzog, W. A. (1982). 'Accommodation of a scientific conception: Toward a theory of conceptual change'. *Science Education,* 66, 211-227.

Piaget, J. (1970). Genetic Epistemology. Columbia University Press, NY.

Rice, M. J. (1972). The case for the Disciplines in the Organization of Social Studies Curricula for Elementary and Secondary Schooling. Paper presented at the Colleges and University Faculty Association of the rational Council for the Social studies (Boston, Massachussets, Nov 21, 1972).

Ryden, B., & Peterson, B. (2020). Foundations of Astrophysics. Cambridge: Cambridge University Press.

Saumure, K., & Given, L. (2008). Convenience sample. In L. Given (Ed.), *The SAGE encyclopedia of qualitative research methods.* (pp. 125-126). Thousand Oaks, CA: SAGE Publications

Schwab, J. J. (1962). The teaching of science as inquiry. In J. J. Schwab & P. Brandwein (Eds.), The teaching of science. Cambridge: Harvard University Press.

Setyadin, A. H.; Suryana, T.G.S.; Utari, S.; Efendi. R.; Liliawati, W.; & Utama, J.A. (2020). Identifying K-10 Students' Learning Difficulties on Learning Kepler's Law using Worksheet: Is It Worth? *J. Phys.: Conf Ser.*, 1467, 012051.

Simon, M.A. (1995). Reconstructing Mathematics Pedagogy from a Constructivist Perspective. Journal for Research in Mathematics Education. 26, 114-145.

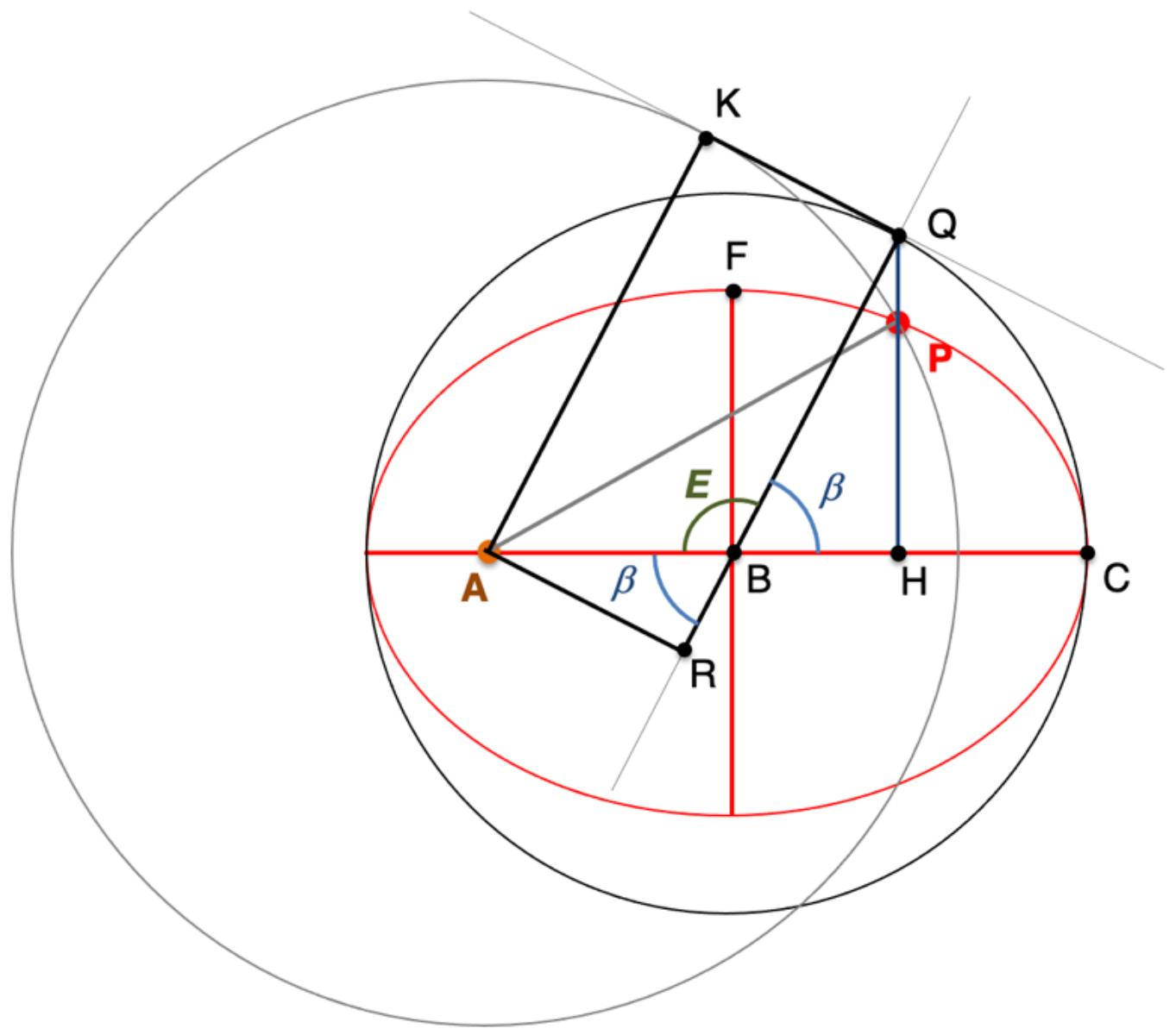

**Fig A1.** Kepler's method to find the curve corresponding to Mars' orbit. The crucial insight was to realize that a perpendicular to BQ at Q is tangent to the circle centered at the Sun (A) and with radius *r* equal to the radius vector that joins the Sun and the planet (P). The eccentricity of Mars' orbit is highly exagerated for clarity.

Waxer M., Morton J.B. (2012) Cognitive Conflict and Learning. In: Seel N.M. (eds) Encyclopedia of the Sciences of Learning. Springer, Boston, MA. https://doi.org/10.1007/978-1-4419-1428-6_280

Welch, W. Klopfer, L., Aikenhead, G. S., & Robinson, J. T. (1981). The role of inquiry in science education: Analysis & recommendations. Science Education, 65, 33-50.

Yu, K.-C.; Denn, G.; & Sahami, K. (2010). Student Ideas about Kepler's Laws and Planetary Orbital Motions. Astronomy Education Review, 9, 010108-1.

## APPENDIX A: Elliptical orbits (Kepler's 1st law)

In this appendix, the instructor guides the demonstration that the shape of the orbit is an ellipse. It works as active learning, with the instructor giving to each student a sheet with the elliptic shape, and a set of drafting tools. This appendix is written in teacher's voice, guiding how to draw Fig A1 step by step. The order: is AP, circle centered at A, PH, circle centered at B, PQ, BQ, *E*, $\beta = H\hat{B}Q$, QK, AK, AKQR, and finally $\beta = A\hat{B}Q$. A video is available at this url.

Having found the orbit, Kepler had no idea what geometrical shape it corresponded to. Yet, Kepler realized, through geometry, some properties this shape had. Consider Fig. A1 (at this point with the only the red curve drawn, points B, C, P, major and minor axes; the focus A can be pre-drawn, or it can be found with the compass, from point F and striking the line of apsides with length BC, the semimajor axis). The Sun is at point A, and Mars at point P. The segment AP, of length *r,* is the radius vector from the Sun to the planet. The orbit is the red curve, which, a priori, we do not know what shape it corresponds to. Some elements are

- The aphelion is point C;
- The line of apsides is bisected at point B, the geometric center of the curve;
- The length of the segment BC is by definition *a*;
- The length of the segment AB is by definition *ae*; *e* is the eccentricity of the ellipse, but we do not know that yet. So far *e* is an *ad hoc* constant, the factor by which we need to move the Sun away from the center.

What we want to find is AP, the radius vector of the orbit (draw AP). Algebraically, one would call it *r* and try to find a mathematical relationship for it. Geometrically, one draw a circle. This circle is centered at A and has radius AP (draw circle). We do not need to find exactly AP; if we find the radius of this circle, at any angle, we find the length of the radius vector. We trace the circle in the hope that a geometric coincidence that helps tell what the length $r$ is becomes obvious.

Another way to find the shape of the curve is to find the coordinates $x$ and $y$ of the planet, and uncover their mathematical relation. We find the coordinate $x$ by drawing the perpendicular from P to the line of apsides, defining point H (draw line and point). The coordinates of the P are $x$=BH and $y$=PH.

Next we define the eccentric anomaly. For that we draw the circumscribed circle, of center B and radius BC=$a$ (draw circle). We prolong the line PH until it intersects the circumscribed circle at point Q (draw PQ). The eccentric anomaly is $E = A\hat{B}Q$ (draw BQ, and the angle $E$). We will also define the auxiliary angle $\beta = 180^o - E = H\hat{B}Q$, the complement of $E$ (draw $\beta$). Given the triangle ΔBHQ, the coordinate $x$ is BQ cos $\beta$. Given BQ=$a$, we found the first coordinate

$$x = a \cos \beta \tag{A1}$$

As for $y$, the triangle ΔAHP can be used. It is a right triangle where AP=$r$ is the hypotenuse; the catheti are PH=$y$, and AH = AB + BH = $ae + x$. Thus,

$$y^2 = r^2 - (ae+x)^2 \tag{A2}$$

Eq (A1) gives the value of $x$, but the value of the radius vector $r$ is so far unknown. Kepler found $r$ in an ingenious way. He realized something curious: *the perpendicular to BQ at Q is tangent to the circle of center A and radius r.* (prolong BQ and draw the perpendicular)

Let K be the tangential point (draw K). Since AK is a radius (draw AK), then AP=AK=$r$. So, if we find AK, we find the value of $r$. Because QK is tangent to the circle, $A\hat{K}Q$ is a right angle. Kepler then prolonged the radius BQ to construct the rectangle AKQR (draw the rectangle and define R). Because this is a rectangle, AK=QR=$r$. The length QR is the sum of the radius (BQ=$a$) and the length BR. This length is given by the right triangle ΔARB. The hypothenuse is AB=$ae$, and the cathetus BR=$ae$ cos $\beta$ (draw $A\hat{B}Q=\beta$). We find thus AP = BQ + BR, or $r = a + ae \cos \beta$. Given $\beta = 180^o - E$

$$r = a(1 - e \cos E) \tag{A3}$$

Having found the radius vector, we substitute Eq. (A3) and Eq. (A1) into Eq. (A2), finding

$$y^2 = a^2 (1-e^2) \sin^2 E \tag{A4}$$

We can then write $\cos^2 E = x^2/a^2$, and $\sin^2 E = y^2/[a^2(1-e^2)]$ and invoke the trigonometric equality $\sin^2 E + \cos^2 E = 1$ to find the relationship between the coordinates

$$\frac{x^2}{a^2} + \frac{y^2}{a^2(1-e^2)} = 1 \tag{A5}$$

This is the equation of an ellipse. The semimajor axis is $a$, and the semiminor axis is $b = a (1-e^2)^{1/2}$.

## APPENDIX B : Short proof of the area of the ellipse

Consider Fig. B1. The main insight is that PH/QH = $b/a$, which is seen because CQ = $a$, and thus QH = $a$ sin $E$, and we have already proven (Eq A4) that PH = $a(1-e^2)^{1/2} \sin E = b \sin E$. The area $A_{\text{Circle}}$ of the circle is 4 times the area of the quadrant. The area of the quadrant can be found by integrating the vertical distances $y_{\text{circ}}$ from $x = 0$ to $x = a$.

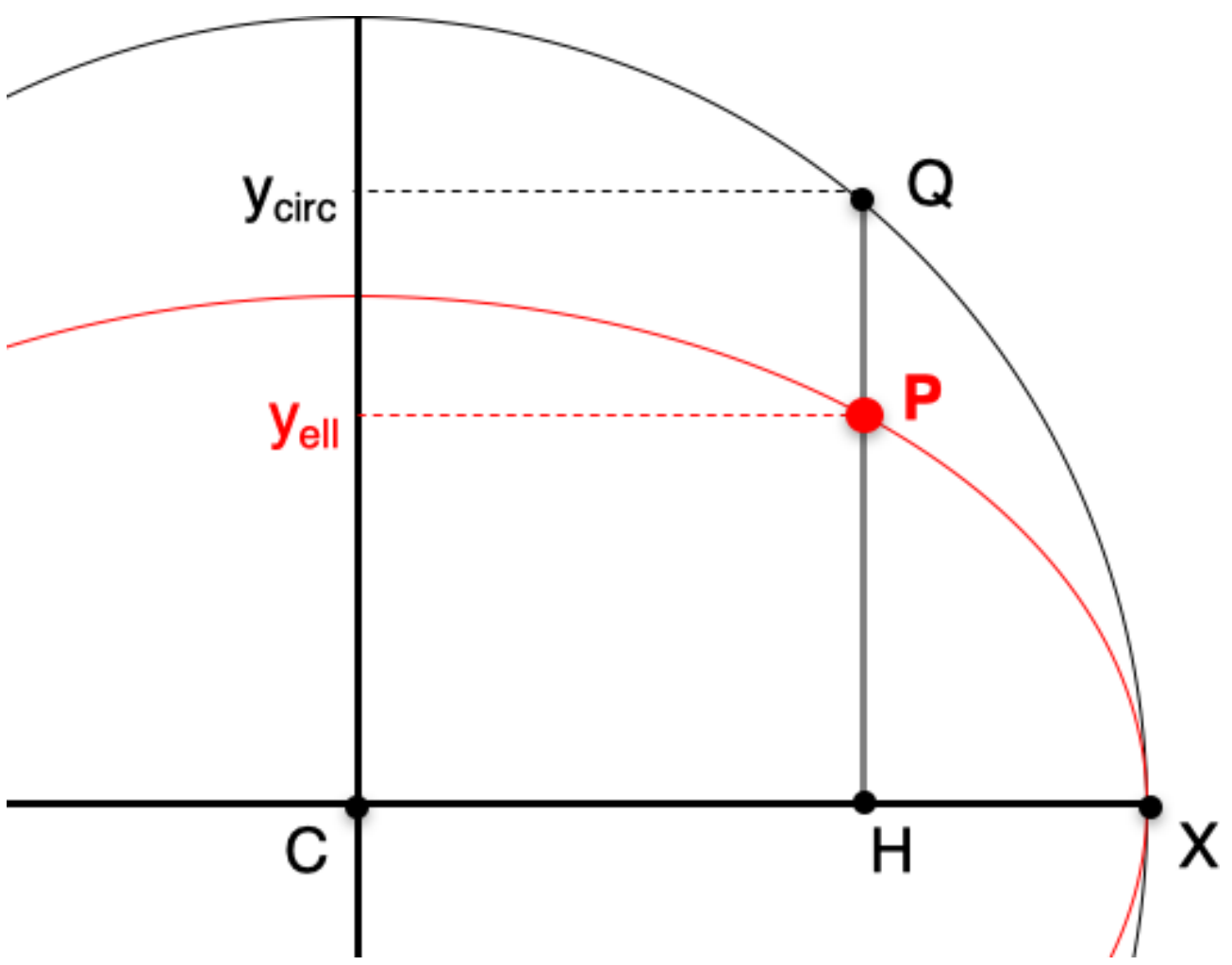

**Figure B1.** The ratio PH/QH = $y_{ell}/y_{circ}$ is equal to $b/a$, where $b$ is the semiminor axis and $a$ the semimajor axis of the ellipse.

$$A_{\text{circle}} = 4\int_0^a y_{\text{circ}}dx \tag{B1}$$

Given $y_{circ} = a \sin E$ and $x = a \cos E$, then $A_{Circle} = \pi a^2$, as expected.

The area of the ellipse is

$$A_{\text{ellipse}} = 4\int_0^a y_{\text{ell}}dx \tag{B2}$$

Because we can write $y_{ell} = b/a \; y_{circ}$, then

$$A_{\text{ellipse}} = 4\frac{b}{a}\int_0^a y_{\text{circ}}dx = \frac{b}{a}A_{\text{circle}} = \pi ab \tag{B3}$$